\documentclass[a4paper]{jpconf}
\usepackage{graphicx,bm}
\usepackage{amsmath,amsfonts}
\newcommand{\mgm}{\ensuremath{\mu}}
\begin{document}
\title{Neutrino spin and spin-flavour oscillations in transversally moving or polarized matter}

\author{Alexander Studenikin}

\address{Department of Theoretical Physics, Faculty of Physics, Lomonosov Moscow State University, Moscow 119991,
Russia\\
Joint Institute for Nuclear Research, Dubna 141980, Moscow
Region,
Russia}
\ead{studenik@srd.sinp.msu.ru}

\begin{abstract}

Studies of an interesting effect of neutrino spin and spin-flavour oscillations engenders by neutrino weak interactions with the transversally moving or polarized matter are reviewed.

\end{abstract}

It is well known that massive neutrinos have nontrivial electromagnetic properties, and at least the magnetic moment is not zero \cite{Fujikawa:1980yx}. Thus, neutrinos do participate also in the electromagnetic interaction (see \cite{Giunti:2014ixa} for a review). The best terrestrial laboratory upper bound on neutrino magnetic moments is obtained by the GEMMA reactor neutrino experiment \cite{Beda:2012zz}. The best astrophysical upper bound was derived from considering stars cooling \cite{Raffelt:1990pj}. The neutrino magnetic moment procession in the transversal magnetic field ${\bf B}_{\perp}$ was first considered in \cite{Cisneros:1970nq}, then spin-flavor precession in vacuum was discussed in \cite{Schechter:1981hw}, the importance of the matter effect was emphasized in \cite{Okun:1986na}. The effect of resonant amplification of neutrino spin oscillations in ${\bf B}_{\perp}$ in the presence of matter was proposed in \cite{Akhmedov:1988uk,Lim:1987tk}, the impact of the longitudinal magnetic field ${\bf B}_{||}$ was discussed in \cite{Akhmedov:1988hd}. Recently we consider in details \cite{Fabbricatore:2016nec} neutrino mixing and oscillations in arbitrary constant magnetic field that have  ${\bf B}_{\perp}$ and ${\bf B}_{||}$ nonzero components in mass and flavour bases. We have also  developed a new (and more precise than the usual one) approach to description of neutrino spin and spin-flavor oscillations in the presence of an arbitrary magnetic field \cite{Dmitriev:2015ega}. Our approach  is based on the use of the stationary states in the magnetic field for classification of neutrino spin states, contrary to the customary approach when the neutrino helicity states are used for this purpose.

In this short note we focus on a very interesting effect in neutrino spin and spin-flavour oscillations in presence of matter background that was first discussed in our paper \cite{Studenikin:2004bu}. In our studies it was shown that neutrino spin and spin-flavour oscillations can be induced not only by the neutrino interaction with a  magnetic field, as it was believed before, but also by neutrino interactions with matter in the case when there is a transversal matter current or matter polarization. In the Conclusions of \cite{Studenikin:2004bu} one finds:

{\it ``The possible emergence of neutrino spin oscillations  owing to neutrino
interaction with matter under the condition that there exists a nonzero transverse current component or
matter polarization is the most important new effect that follows from the investigation
of neutrino-spin oscillations in Section 4. So far, it has been assumed that neutrino-spin oscillations
may arise only in the case where there exists a nonzero transverse magnetic field in the neutrino rest
frame." }

It should be noted that the predicted effect exist regardless of a source of the background matter transversal current or polarization (that can be a background magnetic field, for instance).

Note that the existence of the discussed effect of neutrino spin oscillations engendered by the transversal matter current and matter polarization and its importance for astrophysical applications have been confirmed in a series of recent papers \cite{Cirigliano:2014aoa, Volpe:2015rla, Kartavtsev:2015eva, Dobrynina:2016rwy}.

 Consider, as an example,  an electron neutrino spin procession in the case when neutrinos with the Standard Model interaction are propagating through moving and polarized matter composed of electrons (electron gas) in the presence of an electromagnetic field given by the electromagnetic-field tensor $F_{\mu \nu}=({\bf E}, {\bf B})$. As discussed in \cite{Studenikin:2004bu, Studenikin:2004tv}
(see also \cite{Egorov:1999ah,Lobanov:2001ar, Dvornikov:2002rs}), the generalized Bargmann-Michel-Telegdi equation describes  the evolution of the
three-di\-men\-sio\-nal neutrino spin vector $\vec S $,
\begin{equation}\label{S} {d{\bf S}
\over dt}={2\mgm\over \gamma} \Big[ {\bf S} \times ({\bf
B}_0+{\bf M}_0) \Big],
\end{equation}
where the magnetic field $\bf{B}_0$ in the neutrino rest frame is determined by the transversal
and longitudinal (with respect to the neutrino motion) magnetic and electric field components in the
laboratory frame,
\begin{equation}
{\bf  B}_0=\gamma\Big({\bf B}_{\perp} +{1 \over \gamma} {\bf
B}_{\parallel} + \sqrt{1-\gamma^{-2}} \Big[ {\bf E}_{\perp} \times
\frac{{\bm\beta}}{\beta} \Big]\Big),
\end{equation}
$\gamma = (1-\beta^2)^{-{1 \over 2}}$, $\bm{\beta}$ is the neutrino velocity.

The matter term ${\bf M}_0$ in Eq. (\ref{S}) is also composed of the transversal ${\bf  M}{_{0_{\parallel}}}$
and longitudinal  ${\bf  M}_{0_{\perp}}$ parts,
\begin{equation}
{\bf M}_0=\bf {M}{_{0_{\parallel}}}+{\bf M}_{0_{\perp}},
\label{M_0}
\end{equation}
\begin{equation}
\begin{array}{c}
\displaystyle {\bf M}_{0_{\parallel}}=\gamma{\bm\beta}{n_{0} \over
\sqrt {1- v_{e}^{2}}}\left\{ \rho^{(1)}_{e}\left(1-{{\bf v}_e
{\bm\beta} \over {1- {\gamma^{-2}}}} \right)\right\}-
\displaystyle{\rho^{(2)}_{e}\over {1- {\gamma^{-2}}}} \left\{{\bm\zeta}_{e}{\bm\beta}
\sqrt{1-v^2_e}+ {\left({\bm \zeta}_{e}{{\bf v}_e}\frac{({\bm\beta}{\bf v}_e)}{{1+\sqrt{1-v^2_e}} }\right)}
\right\}, \label{M_0_parallel}
\end{array}
\end{equation}
\begin{equation}\label{M_0_perp}
\begin{array}{c}
\displaystyle {\bf M}_{0_{\perp}}=-\frac{n_{0}}{\sqrt {1-
v_{e}^{2}}}\Bigg\{ {\bf v}_{e_{\perp}}\Big(
\rho^{(1)}_{e}+\rho^{(2)}_{e}\frac
{({\bm\zeta}_{e} {\bf v}_e)} {1+\sqrt{1-v^2_e}}\Big) +
\displaystyle {{\bm\zeta}_{e_{\perp}}}\rho^{(2)}_{e}\sqrt{1-v^2_e}\Bigg\}.
\end{array}
\end{equation}
Here $n_0=n_{e}\sqrt {1-v^{2}_{e}}$ is the invariant number density of
matter given in the reference frame for which the total speed of
matter is zero. The vectors ${\bf v}_e$, and ${\bm \zeta}_e \
(0\leq |{\bm \zeta}_e |^2 \leq 1)$ denote, respectively,
the speed of the reference frame in which the mean momentum of
matter (electrons) is zero, and the mean value of the polarization
vector of the background electrons in the above mentioned
reference frame. The coefficients $\rho^{(1,2)}_e$ calculated
within the extended Standard Model supplied with $SU(2)$-singlet right-handed neutrino
$\nu_{R}$ are respectively,  $\rho^{(1)}_e={\tilde{G}_F \over {2\sqrt{2}\mu }}, \ \ \rho^{(2)}_e =-{G_F \over {2\sqrt{2}\mu}}$,
where $\tilde{G}_{F}={G}_{F}(1+4\sin^2 \theta _W).$

For neutrino evolution between two neutrino states $\nu_{e}^{L}\Leftrightarrow\nu_{e}^{R}$ in presence of the magnetic field and moving matter we get \cite{Studenikin:2004bu} the following equation
\begin{equation}\label{2_evol_eq}
	i\frac{d}{dt} \begin{pmatrix}\nu_{e}^{L} \\ \nu_{e}^{R} \\  \end{pmatrix}={\mgm}
	\begin{pmatrix}
	{1 \over \gamma}\big|{\bf
M}_{0\parallel}+{{\bf B}}_{0\parallel}\big| & \big|{{\bf B}}_{\perp} + {1\over
\gamma}{\bf M}_{0\perp} \big|  \\
	 \big|{{\bf B}}_{\perp} + {1\over
\gamma}{\bf M}_{0\perp} \big| & -{1 \over \gamma}\mid{\bf
M}_{0\parallel}+{{\bf B}}_{0\parallel}\big|  \\
		\end{pmatrix}
	\begin{pmatrix}\nu_{e}^{L} \\ \nu_{e}^{R} \\ \end{pmatrix}.
\end{equation}
Thus, the probability of the neutrino spin oscillations in the adiabatic
approximation is given by (see \cite{Studenikin:2004bu, Studenikin:2004tv})
\begin{equation}\label{ver2}
P_{\nu_L \rightarrow \nu_R} (x)=\sin^{2} 2\theta_\textmd{eff}
\sin^{2}{\pi x \over L_\textmd{eff}},\ \sin^{2} 2\theta_\textmd{eff}={E^2_\textmd{eff} \over
{E^{2}_\textmd{eff}+\Delta^{2}_\textmd{eff}}}, \ \ \
L_\textmd{eff}={2\pi \over
\sqrt{E^{2}_\textmd{eff}+\Delta^{2}_\textmd{eff}}},
\end{equation}
where
\begin{equation}\label{E3}
E_\textmd{eff}=\mgm \big|{{\bf B}}_{\perp} + {1\over
\gamma}{\bf M}_{0\perp} \big|, \ \
\Delta_ \textmd{eff}={\mgm \over \gamma}\big|{\bf
M}_{0\parallel}+{{\bf B}}_{0\parallel} \big|.
\end{equation}

Thus, it follows that even without presence of an electromagnetic field,
${{\bf B}}_{\perp}={{\bf B}}_{0\parallel}=0$,
neutrino spin  oscillations can be induced in the presence of matter
when the transverse matter term ${\bf M}_{0\perp}$ is not zero.
This possibility is realized in the case when the transverse component of the background matter velocity or its transverse polarization is not zero.

The above considerations can be applied to other types of neutrinos and various matter compositions. It is also obvious that for neutrinos with nonzero transition magnetic moments a similar effect of spin-flavour
oscillations exists under the same background conditions.

The possibility of neutrino spin procession and oscillations induced by the transversal matter current or polarization was first discussed in \cite{Studenikin:2004bu, Studenikin:2004tv}. More general case of neutrino spin evolution in the case when neutrino is subjected to general types of non-derative interactions with external scalar $s$, pseoudoscalar $\pi$, vector $V_{\mu}$, axial-vector $A_{\mu}$, tensor $T_{\mu\nu}$ and pseudotensor $\Pi_{\mu\nu}$ fields was considered in \cite{Dvornikov:2002rs}. From the obtained general neutrino spin evolution equation it follows that neither scalar $s$ nor pseudoscalar $\pi$ nor vector $V_{\mu}$ fields can induce neutrino spin evolution. On the contrary, within the general consideration of neutrino spin evolution it was shown that electromagnetic (tensor) and weak (axial-vector) interactions can contribute to the neutrino spin evolution.

The author is thankful to Konstantin Kouzakov for discussions. The author is also thankful to the organizers of the  27th International Conference on Neutrino Physics and astrophysics for the invitation to attend this very interesting event. This work is supported by the Russian Basic Research Foundation grants No. 14-22-03043, 15-52-53112 and 16-02-01023.

\end{document}